\documentstyle[manuscript,aps,prbbib]{revtex}

\tighten
\newcommand{\be}{\begin{equation}}
\newcommand{\ee}{\end{equation}}
\newcommand{\bea}{\begin{eqnarray}}
\newcommand{\eea}{\end{eqnarray}}
\newcommand{\bd}{\begin{displaymath}}
\newcommand{\ed}{\end{displaymath}}
\newcommand{\bi}{\begin{itemize}}
\newcommand{\ei}{\end{itemize}}
\newcommand{\bc}{\begin{center}}
\newcommand{\ec}{\end{center}}
\newcommand{\bfl}{\begin{flushleft}}
\newcommand{\efl}{\end{flushleft}}
\newcommand{\bfr}{\begin{flushright}}
\newcommand{\efr}{\end{flushright}}
\newcommand{\f}{\frac}

\newcommand{\au}{\u{a}}

\newcommand{\s}{\c{s}}

\def\6{\partial} \def\a{\alpha} \def\b{\beta}
\def\g{\gamma} \def\d{\delta}  \def\e{\epsilon}
\def\z{\zeta} \def\h{\eta} \def\th{\theta}
  \def\l{\lambda}
\def\m{\mu} \def\n{\nu} \def\x{\xi} 
\def\r{\rho} \def\ss{\sigma} 

\def\o{\omega} \def\G{\Gamma}

\def\={\!\!\!&=&\!\!\!}
\def\+{\!\!\!&&\!\!\!+~}
\def\-{\!\!\!&&\!\!\!-~}

\newcommand{\DD}{{\cal D}}

\newcommand{\FF}{{\cal F}}

\newcommand{\TT}{{\cal T}}

\def\nn{\nabla}
\def\ne{\nabla_{\eta}}
\def\nx{\nabla_{\xi}}
\def\nse{\nabla^{S}_{\eta}}
\def\nsx{\nabla^{S}_{\xi}}
\def\Rex{R(\eta ,\xi )}
\def\Rsex{R^S (\eta ,\xi )}

\begin{document}

\title{Constraints on spacetime manifold
in Euclidean supergravity in terms of Dirac eigenvalues}
\author{C. Ciuhu and Ion V. Vancea}
\address{Department of Theoretical Physics\\
Babes-Bolyai University of Cluj\\
Str. M. Kogalniceanu Nr.1, RO-3400 Cluj, Romania}
\date{\today}
\maketitle

\begin{abstract}
We generalize previous work on Dirac eigenvalues as dynamical variables
of Euclidean supergravity. The most general set of constraints on the
curvatures of the tangent bundle and on the spinor bundle of the spacetime
manifold, under which spacetime addmits Dirac eigenvalues as observables,
are derived.
\end{abstract}

\section{Introduction}

The formulation of gravity in terms of noncommutative
geometry aims at giving a resolution to the problem of quantum gravity.
There are already several approaches to the subject, for instance
Connes-Dixmier-Wodzicki formulation \cite{ac} based on Dixmier trace
\cite{dix} and Wodzicki residue formula \cite{wod} which was
generalized to any dimensional gravity \cite{kw}, the formulation in
terms of linear connections \cite{cffg} or the theory based on the
spectral action principle \cite{cc} which also provides the action of the
Standard Model \cite{mv}. All of these theories are formulated on Euclidean
spacetime manifolds because the spectral triple necessary to formulate
the noncommutative geometry can be defined only in this case. However,
there are attempts to formulate spectral triples with Minkowskian signature
\cite{mst}. For fundamental and recent reviews of these topics the
reader should see \cite{rev}. A supersymmetric extension of the action
that obeys the spectral priciple has been recently reported in \cite{ahc}
while in \cite{aa} has been shown that quantum versions of Riemann
objects exhibit a noncommutative structure.

In a very recent paper \cite{lr} one argues that Dirac eigenvalues are
the observables of Euclidean gravity. In a previous work one of us developed
the idea that Dirac eigenvalues can play the same role in the case
of Euclidean supergravity once some constraints are imposed on the covariant
phase space as well as on Dirac eigenspinors \cite{viv1}. Moreover, it has been
shown that a spacetime manifold which supports such of a description of
supergravity should also satisfy some constraints. Indeed, by applying the
parallel transport to the constraints on the phase space and under
some simplifying assumptions, a set of equations that should be
satisfyied by the curvatures of the tangent bundle and of the spinor
bundle, respectively, were derived \cite{viv2}.

Here, we extend our understanding of the compatibility between Dirac
eigenvalues as observables 
of Euclidean supergravity 
and the geometrical structure of spacetime, by
investigating the most general set of constraints on the two curvatures
resulting from the parallel transport of the constraints on the phase space,
or primary constraints, as well as from the parallel transport  of
the constraints on Dirac eigenspinors, or secondary constraints.

The plan of the paper is as follows. In Sec. II we review the basic results
obtained in the case of Euclidean supergravity and present the basic ideas
which allow us to derive the constraints on the curvatures. In Sec. III we
determine these constraints in the most general form. In Sec. IV we summarize
the results and make some concluding remarks. Units have been chosen so that
$8\pi G =1$.

\section{Primary and secondary constraints of Euclidean Supergravity}

The aim of this section is the brief presentation of the constraints of
Euclidean supergravity that should be imposed locally in order to have Dirac
eigenvalues as observables of the system. For extended discussions the reader
should see \cite{viv1,viv2}.

Consider a compact four dimensional (4D) (spin)manifold without boundary $M$,
on which Euclidean minimal supergravity is defined. The $N=1$ on-shell
supergraviton multiplet contains the graviton, represented by vierbein
fields $e^{a}_{\m}(x)$, where $ \m = 1, \ldots ,4$ are spacetime indices and
$a=1, \ldots , 4$ are internal Euclidean indices, and the gravitino fields
$\psi^{\a}_{\m}(x)$, which are required to satisfy, instead of the usual Majorana
condition, the following one $\bar{\psi} = \psi^{T}C$ \cite{spin}. This
replacement is necessary because the group of local rotations $SO(4)$ does
not admitt a Majorana representation. Alternatively, one could work with
symplectic spinors which could also be more suitable for generalizations
to higher dimensional Euclidean supergravity \cite{avp}. The covariant
phase space of the theory is defined to be the phase space of the solutions
of the equations of motion modulo the gauge transformations, which are
4D diffeomorphisms, local $SO(4)$ rotations and $N=1$ 
local supersymmetry. The
observables of the theory are functions on the phase space. On $M$, the Dirac
operator is defined in the presence of local supersymmetry as
\be
D=\overcirc{D} + K, \label{diracop}
\ee
where
\be
\overcirc{D} = i\g^a e_{a}^{\m} (\6_{\m} + 
\frac{1}{2}\overcirc{\o}_{\m bc}(e) \ss^{bc} ), \label{diracopo}
\ee
is the Dirac operator in the absence of local supersymmetry and
\be
K=i\g^a e^{\m}_{a}K_{\m bc}(\psi ) \ss^{bc}.  \label{bkap}
\ee
Here, $\g^a$'s form an Euclidean representation of the Clifford algebra
$C_4$, i. e. $\{ \g^a, \g^b \} = 2\delta^{ab}$, $\overcirc{\o}_{\m bc}(e)$
is the usual spin connection in the absence of supersymmetry and
$\ss^{bc} = \frac{1}{4}[\g^a , \g^b ]$. The contortion term, required by 
for local supersymmetry, is given by
\be
K_{\m ab}(\psi) = \f{i}{4}(\bar{\psi_{\m}}\g_a \psi_{b} -
\bar{\psi_{\m}}\g_{b}\psi_a + \bar{\psi_b}\g_{\m}\psi_a ). \label{cont}
\ee
Since we have assumed that $M$ is compact, the Dirac operator which is an
elliptic operator, has a discrete spectrum and a complete set of eigenspinors
\be
D \chi^n = \l^n \chi^n ,\label{diraceq}
\ee
where $ n=0,1,2, \ldots$. $\l^n$'s depend on $(e,\psi )$ and thus define a
discrete family of functions on the space of all supermultiplets. 
If we denote the space of all gravitational supermultiplets by $\FF$, then
$\l^n $'s define a discrete family of functions on $\FF$ since these
functions depend on $(e, \psi )$. This is a consequence of the dependence
of $D$ on $(e,\psi )$.
Therefore,
$\l^n $'s define a discrete family of real valued functions
on $\FF$ and a function from $\FF$ into the space of infinite sequences 
$R^{\infty}$
\bea
\l^{n} &:& \FF \longrightarrow R ~~~~,~~(e,\psi ) \rightarrow
\l^{n}(e,\psi ) \\ \label{functs}
\l^{n} &:& \FF \longrightarrow R^{\infty}~~,~~(e,\psi ) \rightarrow
\{ \l^{n}(e,\psi ) \}. \label{seqs}
\eea
Let us note that, in general, Dirac eigenvalues are not invariant under the
gauge transformations which represents a different situation from the gravity case.
Therefore, they cannot be immediately use as observables and we must identify
the circumstances which allow us to treat $\l^n$'s as observables.

The basic requirement is that Dirac eigenvalues be gauge invariant. The
action of gauge transformations on the supergraviton are given by the following
equations
\bea
\d e^{a}_{\m} &=& \xi^{\n} \6_{\n} e^{a}_{\m} + \th^{ab} e_{b\m}
+ \frac{1}{2}\bar{\e}\g^{a}\psi_{\m}\\
\d \psi^{\a}_{\m} &=& \xi^{\n} \6_{\n} \psi^{\a}_{\m} +
\th^{ab} (\ss_{ab})^{\a}_{\b} \psi_{\m}^{\b} +
\DD_{\m}\e^{\a} 
\label{gaugetr}
\eea 
where $\xi = \xi^{\nu}\6_{\n}$ is an infinitezimal vector field on $M$,
$\th_{ab}=-\th_{ba}$ parametrize an infinitezimal rotation and $\epsilon$
is an infinitezimal "Majorana" spinor field. Here, $\DD_{\m}$ is the nonminimal
covariant derivative acting on spinors and associated to the minimal one
acting on vectors and on graviton according to the usual rules
\bea
\DD_{\m}A^{\n}  & = & \6_{\m}A^{\n} + \G^{\n}_{\m \ss}A^{\ss} \\
\DD_{\m } A^a   & = & \6_{\m}A^{a} + \o_{\m b}^{a}A^{b} \\
\DD_{\n}e^{a}_{\m } & = & \6_{\n} e^{a}_{\m } - \G^{\r}_{\m \n}e^{a}_{\r}
                        + \o^{a}_{\n b} e^b_{\m } , \label{covderiv}
\eea
where $A=A^{\m }\6_{\m }$ is an arbitrary spacetime vector, 
$A=A^{a }\6_{a}$ is an arbitrary $SO(4)$ vector and
$\G^{\r}_{\m \n}$'s are the Christoffel symbols.
Then the nonminimal
covariant derivative is given by
\be
\DD_{\m} \phi =
\6_{\m} \phi +\o_{\m ab} \ss^{ab}
\phi
\label{spinder}
\ee
for an arbitrary spinor $\phi $. Under local supersymmetry ${\o}_{\m}^{ab}$
 transforms as
\be
\d \o_{\m}^{ab} = A_{\m}^{ab} - \frac{1}{2}e_{\m}^{b}A_{c}^{ac}
+\frac{1}{2}e_{\m}^{a}A_{c}^{bc} , \label{spinsp}
\ee
where
\be
A_{a}^{\m \n} = \bar{\e}\g_{5}\g_{a}\DD_{\l}\psi_{\r}\e^{\n \m \l \r},
\label{coefsp}
\ee
while under an $SO(4)$ rotations the coefficients of spin
connection transform as
\be
\delta \omega_{\m ab} = i [ \th \ss , \o_{\m ab} ] - i \6_{\m}\th \ss M_{ab}.
\label{rotsp}
\ee

As was shown in \cite{viv1}, the variations of $\l^n$'s under gauge
transformations vanish only if the following relations hold:
\be
\TT^{n\m }_{a} \6_{\n}e^{a}_{\m}
- \G^{n \m }_{\a}\6_{\n}\psi_{\m}^{\a} =0 \label{firstc}
\ee
as a consequence of diffeomorphisms, 
\be
\TT^{n\m }_{a}e_{b\m} + \G^{n\m }\ss_{ab}\psi_{\m}=0 \label{secondc}
\ee
as a consequence of local $SO(4)$ transformations and
\be
\TT^{n\m }_{a}\bar{\e}\g^{a}\psi_{\m} + \G^{n\m }\DD_{\m}\e =0
\label{thirdc}
\ee
which follows from  $N=1$ local supersymmetry. The less obvious terms
in Eqs. (\ref{firstc}), (\ref{secondc}) and (\ref{thirdc}) are given
by
\be
\TT^{n \m}_{a} (x) =
\frac{\d \l^n }{\d e^{a}_{\m}}
=\langle \chi^n | \frac{\d D}{\d e^{a}_{\m}} \chi^n \rangle ~~,~~
\G^{n \m }_{\a} =
\frac{\d \l^n}{\d \psi^{\a}_{\m}} =
\langle \chi^n | \frac{\d D}{\d \psi^{\a}_{\m}} \chi^n \rangle,
\label{coef}
\ee
where the scalar product is defined in the Hilbert
space of spinors on $M$ and is given by
\be
\langle \chi , \phi \rangle = \int \sqrt{e} \chi^* \phi \ .
\ee

Now once the Eqs. (\ref{firstc}), (\ref{secondc}) and (\ref{thirdc})
(called primary constraints) hold, there is another set of constraints
(called secondary constraints) which should be imposed on the Dirac
eigenspinors. The reason for that is simply the fact that Dirac eigenvalue
problem (\ref{diraceq}) also transforms under (\ref{gaugetr}) and in
its variations we must take into account the fact that the variations of
eigenvalues vanish if they are to be considered observables. The equations
reflecting this consistency condition are the following ones:
\be
\{ [b^{\m}(\x ) - c(\l ,\x )^{\m} ]\6_{\m} + f(\x ) \} \chi^n = 0,
\label{firstcc}
\ee
which folows from (\ref{firstc}), where we have employed the following
notations
\bea
b^{\m}(\x )  &=& i \g^{a} b_{a}^{\m}(\x )~~,~~
b_{a}^{\m}(\x ) = \x^{\n}\6_{\n}e_{a}^{\m} -e_{a}^{\n}\6_{\n}\x^{\m}
-2e_{a}^{\n}\x^{\m}\o_{\n bc}\ss^{bc}\\  
c(\l ,\x )^{\m} &=& (\l^n - D)\x^{\m}~~,~~   
f(\x ) = i\g^{a}\x^{\n}\6_{\n}(e_{a}^{\m}\o_{\m bc})\ss^{bc}. \\
c(\l ,\x )^{\m} &=& (\l^n - D)\x^{\m}. 
\eea
From the second of the primaries, namely (\ref{secondc}), 
the following equation results
\be
[\th_{a}^{a}D -g(\th ) +h(\th ) ]\chi^n = 0.
\label{secondcc}
\ee
where
\bea
g(\th ) &=& [\g^c e_{c}^{\m}([{\bf \th \ss},\o_{\m ab}] -
\6_{\m}{\bf \th \ss }M_{ab})]\ss^{ab} \\
h(\th ) &=&  i (\l^n - D) {\bf \th \ss}. 
\eea
Finally, $N=1$ local supersymmetry implies
\be
[j^{\m}_{a} (\e ) \6_{\m} + k_{a} (\e ) +l_{a} ]\chi^n =0 \label{thirdcc}
\ee
where
\bea
j^{\m }_{a} (\e ) & = & \frac{1}{2}\g_{a}\bar{\e}\psi^{\m} ~,
~~~k_{a}(\e ) = \frac{1}{2}\g_{a}\bar{\e}\psi^{\m}\o_{\m cd}\ss^{cd} \\
l_a &=&e_{a}^{\m}[A_{\m cd} - \frac{1}{2}e_{\m d}A_{ec}^{e} +
\frac{1}{2}e_{\m c}A_{ed}^{e}]\ss^{cd}.
\eea

Some comments are now in order. The first set of constraints given by
eqs. (\ref{firstc}), (\ref{secondc}) and (\ref{thirdc}) should be interpreted
as constraints on the phase space of the theory. Therefore, the observables of
the theory are defined on the intersection of the solutions of the equations
of motion with the solutions of the primary constraints. This intersection
represents a subset of the set of all supergravitons. The second set of
constraints given by eqs. (\ref{firstcc}), (\ref{secondcc}) and (\ref{thirdcc})
points out the set of Dirac eigenspinors for which the variations of the
correponding eigenvalues under gauge transformations vanishes. This
shows us a second restrictions, this time on the possible observables of the
theory.

We must notice that the entire discussion has a local character up to now.
For
the theory to be fully consistent, we should address the problem of
compatibility of the primary and secondary constraints with the global
structure of spacetime manifold $M$. A very general and important class of
manifolds can be obtained by considering the parallel transport of the
two types of constraints from one point of $M$ to another one
along two different arbitrary paths. The compatibility of the geometric
structure of $M$ with the two sets of constraints can then be rephrased as
the condition of obtaining the same result after the transport of any of
the constraints along the two paths. This problem was partially
addressed in \cite{viv2} where a set of constraints on the curvatures of the
tangent bundle and spinor bundle, respectively, was obtained. However,
the equations obtained in \cite{viv2} were calculated under two restrictions:
firstly, there were analysed only the displacements of the primaries.
Secondly, there was made the simplfying assumption that Dirac eigenspinors
are subject to parallel transport along the two paths.

In the next section we obtain the constraints on the manifold $M$ by relaxing
the above assumption, that is by considering an arbitrary
transformation on the Dirac eigenspinors. Moreover, the constraints arising
from the transport of the secondaries will also be deduced there in the
general
case. The technique used here is the same as the one employed in
\cite{viv2}, but to make the computations in the following section more
transparent, we give a brief account of it before ending this paragraph.

Let us consider two congruences $c(\l )$ and $d(\m )$ on $M$ with $\l$ and
$\m $ the parameters of the curves, and let us take a curvilinear rectangle
at the intersection of the two congruences, $\{ Q, P, R, S \} \in
c(\l ) \cap d(\m )$, where $|QP| \in c, |RS| \in c, |PR| \in d, |QS| \in d$
and the vertical bars denote the curvilinear segment. Assume that the
lengths of the sides of the rectangle are $\l$ and $\m$ measured in units
of natural parameters of $c(\l )$ and $d(\m )$, respectively. We take two
commuting vector fields $\xi$ and $\eta$ defined along $c(\l )$ and
$d(\m )$, respectively and we denote the path $Q\rightarrow P\rightarrow R$
by 1 and the path $Q\rightarrow S\rightarrow R$ by 2. Any object carrying the
subscript 1 or 2 will be understood as transported along the respective path.
Now, since bosons as well as fermions enter the objects to  be transported
along the two paths, we need two connections, $\nabla$  on the
tangent bundle and $\nabla^S$ on the spinor bundle, related to the minimal
covariant derivative and to the nonminimal covariant derivative, respectively.

Then a boson transported along path 1 is given by
\be
A_1 = e^{\m \nabla_{\eta} } e^{\l \nabla_{\xi} }A .
\label{transpbos}
\ee
If we consider $\l$ and $\m$ small and power expand (\ref{transpbos}) up to
the second order, we obtain
\be
A_1 = A^{(0)}_1 + \m A^{(1)}_1 + \l A^{(2)}_1 + \f{\m^2}{2} A^{(3)}_1
    + \f{\l^2}{2} A^{(4)}_1 + \m \l A^{(5)} . \label{powers}
\ee
Since the equality between two objects transported along path 1 and path 2
should hold at all orders in power expansion (\ref{powers}), we expect to
obtain some information about the two curvatures from the coefficient
of $\m \l $. Notice that for products of two objects transported along the two
paths the following relation holds:
\be
A_1 B_1 - A_2 B_2|_{\m \l } = A^0 (B^{(5)}_{1NS} -B^{(5)}_{2NS})+
(A^{(5)}_{1NS} -A^{(5)}_{2NS})B^0 , \label{relhold}
\ee
where $NS$ subscript denotes the nonsymmetric part with respect to $\ne \nx $. 
Equation (\ref{relhold}) can be easily generalized to an arbitrary number of
terms entering the product.

We conclude this section by noting that the same
considerations hold true for the case of fermions transported along path 1 and
path 2, with $\nabla$ replaced by $\nabla^S$. For other details the reader
should see Sec. V and Appendix B and C of \cite{viv2}.

\section{Compatibility of primary and secondary constraints with
spacetime geometry}

In this sections we are going to compute the constraints
on the curvatures $R$ and $R^S$ of the tangent bundle and spinor
bundle of the manifold $M$. We notice that, since the two vector
fields $\xi$ and $\eta$ commute, the two curvatures can be
expressed as:
\be
\Rex = [ \ne ,\nx ] ~~~,~~~ \Rsex = [ \nse , \nsx ]. \label{curvats}
\ee

Let us see what happens when the first primary (\ref{firstc}) is
transported along the two paths. The general strategy is to transport each
of the fields, e. g. $e^a_{\m}, \psi^{\a}_{\m}$, etc. from one point to
another and then to reconstruct the whole equation. These terms obey
the Eq. (\ref{transpbos}) with the power expansion truncated up to
the second order given by Eq. (\ref{powers}). Then from such of transported
objects we form differences like in Eq. (\ref{relhold}) which must cancel
because we require that the result of the parallel transport be independent
of the path. In deducing the corresponding relation in \cite{viv2} one assumed
that the Dirac eigenspinors $\chi^n$'s are subject to parallel transport.
That implies, of course, that certain simplifications occur.

Consider now that the Dirac eigenspinors transform along path 1 in a general
way, i.e.
\be
\chi_1 = e^{\m \nabla^{S}_{\eta} } e^{\l \nabla^{S}_{\xi} }\chi .
\label{diractrans}
\ee
and in a similar manner along the path 2. From the power expansion
(\ref{powers}) we can see that
\be
\chi^{n5}_{2} - \chi^{n5}_{1} = \Rsex\chi^n .
\label{fivedirac}
\ee
After some agebra we can extract the coefficient of $\m \l$ from the power
expansion. It can be shown, exactly as in \cite{viv2},
that the rest of the terms vanish. Therefore, we obtain the following
equation:
\bea
( \langle \Rsex \chi^n | \TT^{\m}_{a} | \chi^n \rangle +
\langle \chi^n | \TT^{\m}_a | \Rsex \chi^n \rangle ) \6_{\n}e^{a}_{\m} +
\nonumber\\
( \langle \Rsex \chi^n | \G^{\m}_{\a} | \chi^n \rangle +
\langle \chi^n | \G^{\m}_{\a} | \Rsex \chi^{n} \rangle ) \6_{\n}\psi^{\a}_{\m}
+ \nonumber\\
\langle \chi^n |i \g^d \d_{da} \d^{\m \r}
(\6_{\r} + \f{1}{2}\overcirc{\o}^{(0)}_{\r fg}\ss^{fg}) -
\nonumber\\
\f{1}{8}\g_a g^{\m \n} \sum_{(\n ,b,c)}
[\bar{\psi}_{\n} \g_b \psi_c ]
\ss^{bc}|\chi^n \rangle
[\Rex (\6_{\n})e^{a}_{\m})] +
\nonumber\\
\langle \chi^n | i \g^d \d_{da} \d^{\m \n}
[(\Rex \6_{\n} ) +
\f{1}{2}\overcirc{\o}^{(5)}_{\n fgNS}(\Rex )\ss^{fg}]
+
\nonumber\\
i\g^d[\f{\d}{\d e^{a}_{\m}}(\Rex e^{\r}_d )
(\6_{\r} + \f{i}{2}\overcirc{\o}^{(0)}_{\r fg}\ss^{fg}) -
\f{\d}{\d e^{a}_{\m}}[(\Rex e_{\n d}] \d^{\n \r}
(\6_{\r} + \f{1}{2}\overcirc{\o}^{(0)}_{\r fg}\ss^{fg})] -
\nonumber\\
\f{1}{8}\g_a g^{\m \r}\sum_{(\r ,b,c)}
[\Rsex (\bar{\psi_{\r}}\g_b \psi_c) ]
\ss^{bc}|\chi^n \rangle
\6_{\n} e^{a}_{\m} +
\nonumber\\
\f{1}{8}\langle \chi^n | \g^a e^{\r}_{a} \f{\d }{\d \psi^{\a}_{\m}} 
[\sum_{(\r ,b,c)}
[\bar{\psi}_{\r}\g_b \psi_c ]]\ss^{bc}|\chi^n \rangle
[\Rsex (\6_{\n} \psi^{\a}_{\m} )]-
\nonumber\\
\langle \chi^n |i \g^a \{ e^{\r}_{a} \f{\d}{\d \psi^{\a}_{\m}}
[\f{i}{8}\sum_{(\r ,b,c)}[\Rsex (\bar{\psi}_\r \g_b \psi_c)]+
[\d^{\m}_{\n} \d^{\b }_{\a } (\Rex e^{\r}_{a})-
\nonumber\\
\f{1}{8}\f{\d}{\d \psi^{\a }_{\m }} 
( \Rsex \psi^{\b }_{\n }) e^{\r }_{a}] 
\f{\d}{\d \psi^{\b}_{\n}}[\sum_{(\r ,b,c)}[\bar{\psi}_{\r}\g_b \psi_c ]]
\} \ss^{bc} |\chi^n \rangle \6_{\n} \psi^{\a}_{\m} = 0,
\label{constf}
\eea
Eq. (\ref{constf}) represent the
results of transporting the first primary along the two paths together
with the condition of vanishing the coefficient of $\m \l $. In a similar
manner we can derive the equation resulting from the transport of the
second primary which reads, after some lenghty calculations, as follows
\bea
(\langle \Rsex \chi^n | \TT^{\m}_{a} | \chi^n \rangle +
\langle \chi^n | \TT^{\m}_a | \Rsex \chi^n \rangle )e_{b\m} +
\nonumber\\
(\langle \Rsex \chi^n | \G^{\m}_{\a} | \chi^n \rangle +
\langle \chi^n | \G^{\m}_{\a} | \Rsex \chi^n \rangle )
(\ss_{ab})^{\a}_{\b}\psi^{\b}_{\m}
+\nonumber\\
\langle \chi^n |i \g^d \d_{da} \d^{\m \n}
(\6_{\n} + \f{1}{2}\overcirc{\o}^{(0)}_{\n fg}\ss^{fg}) -
\nonumber\\
\f{1}{8} \g_a g^{\m \n} \sum_{(\n ,d,c)}
[\bar{\psi}_{\n} \g_d \psi_c ]
\ss^{dc}|\chi^n \rangle
(\Rex e_{b \m}) +
\langle \chi^n | i \g^d \d_{da} \d^{\m \n}
[(\Rex \6_{\n} ) +
\nonumber\\
\f{1}{2}\overcirc{\o}^{(5)}_{\n fgNS}(\Rex )\ss^{fg}]
+i\g^d[\f{\d}{\d e^{a}_{\m}}(\Rex e^{\r}_d )
(\6_{\r} + \f{1}{2}\overcirc{\o}^{(0)}_{\r fg}\ss^{fg}) -
\nonumber\\
\f{\d}{\d e^{a}_{\m}}[(\Rex e_{\n d}] \d^{\n \r}
(\6_{\r} + \f{1}{2}\overcirc{\o}^{(0)}_{\r fg}\ss^{fg})] -
\nonumber\\
\f{1}{8}\g_a g^{\m \n}\sum_{(\n ,d,c)}
[\Rsex (\bar{\psi_{\n}} \g_d \psi_c )]
\ss^{dc}|\chi^n \rangle e_{b \m} +
\nonumber\\
\langle \chi^n | \g^d e^{\r}_{d} \f{\d }{\d \psi^{\a}_{\m}} 
\f{1}{2}[\sum_{(\r ,t,c)}
[\bar{\psi}_{\r}\g_t \psi_c ]]\ss^{tc}|\chi^n>
[\ss_{ab}]^{\a}_{\g }(\Rsex \psi^{\g}_{\m}) 
+
\nonumber\\
<\chi^n |i\g^d \{ e^{\r}_{d} \f{\d}{\d \psi^{\a}_{\m}}
[\f{1}{2}\sum_{(\r ,t,c)}[\Rsex (\bar{\psi}_\r \g_t \psi_c)]+
[\d^{\m}_{\n} \d^{\b }_{\a } (\Rex e^{\r}_{d})+
\nonumber\\
\f{1}{2}\f{\d}{\d \psi^{\a }_{\m }} 
( \Rsex \psi^{\b }_{\n }) e^{\r }_{d}] 
\f{\d}{\d \psi^{\b}_{\n}}[\sum_{(\r ,t,c)}[\bar{\psi}_{\r}\g_t \psi_c ]]
\} \ss^{tc} |\chi^n \rangle [\ss_{ab}]^{\a}_{\g}\psi^{\g}_{\m} = 0
\label{consts} .
\eea
If we consider now the third primary and transport it along the two paths,
we obtain the following result
\bea
(\langle \Rsex \chi^n | \TT^{\m}_{a} | \chi^n \rangle +
\langle \chi^n | \TT^{\m}_a | \Rsex \chi^n \rangle )
\bar{\epsilon}\g^a \psi_{\m} +
\nonumber\\
(\langle \Rsex \chi^n | \G^{\m}_{\a} | \chi^n \rangle +
\langle \chi^n | \G^{\m}_{\a} | \Rsex \chi^n \rangle )
\DD_{\m}\epsilon^{\a}
+\nonumber\\
\langle \chi^n |i \g^d \d_{da} \d^{\m \n}
(\6_{\n} + \f{1}{2}\overcirc{\o}^{(0)}_{\n fg}\ss^{fg}) -
\nonumber\\
\f{1}{8} \g_a g^{\m \n} \sum_{(\n ,b,c)}
[\bar{\psi}_{\n} \g_b \psi_c ]
\ss^{bc}|\chi^n \rangle 
[(\Rsex \bar{\e})\g^a \psi_{\m} +
\bar{\e}\g^a (\Rsex \psi_{\m})] +
\nonumber\\
\langle \chi^n | i \g^d \d_{da} \d^{\m \n}
[(\Rex \6_{\n} ) +
\f{1}{2}\overcirc{\o}^{(5)}_{\n fgNS}(\Rex )\ss^{fg}]
\nonumber\\
+i\g^d[\f{\d}{\d e^{a}_{\m}}(\Rex e^{\r}_d )
(\6_{\r} +\f{1}{2} \overcirc{\o}^{(0)}_{\r fg}\ss^{fg}) -
\f{\d}{\d e^{a}_{\m}}[(\Rex e_{\n d}] \d^{\n \r}
(\6_{\r} + \f{1}{2}\overcirc{\o}^{(0)}_{\r fg}\ss^{fg})] -
\nonumber\\
\f{1}{8}\g_a g^{\m \n}\sum_{(\n ,b,c)}
[\bar{\psi}_{\n}\g_b (\Rsex \psi_c ) + (\Rsex \bar{\psi_{\n}})\g_b \psi_c ]
\ss^{bc}|\chi^n \rangle \bar{\e}\g^a \psi_{\m} +
\nonumber\\
\f{1}{8}\langle \chi^n |i \g^a e^{\r}_{a} \f{\d }{\d \psi^{\a}_{\m}} 
[\sum_{(\r ,b,c)}
[\bar{\psi}_{\r}\g_b \psi_c ]]\ss^{bc}|\chi^n \rangle
[(\Rex \6_{\m} )\e^{\a} +
\nonumber\\
\6_{\m}(\Rsex \e^{\a})+
\f{1}{2}[\overcirc{\o}^{(5)}_{\m abNS}(\Rex )[\ss^{ab}]^{\a}_{\b}
\e^{\b}+
\nonumber
\overcirc{\o}^{(0)}_{\m ab}[\ss^{ab}]^{\a}_{\b}
(\Rsex \e^{\b})]+ 
\nonumber\\
\f{1}{8}[\sum_{(\m ,b,c)}[\bar{\psi}_{\m}\g_b \psi_c ][\ss^{bc}]^{\a}_{\b}
(\Rsex \e^{b}) + \sum_{(\m ,b,c)}[\bar{\psi}_{\m}\g_b (\Rsex \psi_c )+
\nonumber\\
(\Rsex \bar{\psi}_{\m})\g_b \psi_c ][\ss^{bc}]^{\a}_{\b}\e^{\b}]+
\langle \chi^n |i\g^a \{ e^{\r}_{a} \f{\d}{\d \psi^{\a}_{\m}}
[\f{1}{8}\sum_{(\r ,b,c)}[(\Rsex \bar{\psi}_\r )\g_b \psi_c +
\nonumber\\
\bar{\psi}_\r \g_b (\Rsex \psi_c )] +
[\d^{\m}_{\n} \d^{\b }_{\a } (\Rex e^{\r}_{a})+
\nonumber\\
\f{\d }{\d \psi^{\a }_{\m }} 
( \Rsex \psi^{\b }_{\n }) e^{\r }_{a}] \times
\nonumber\\
\f{\d }{\d \psi^{\b}_{\n}}[\f{1}{8}
\sum_{(\r ,b,c)}[\bar{\psi}_{\r}\g_b \psi_c ]]
\} \ss^{bc} |\chi^n \rangle \DD_{\m}\e^{\a} = 0 .
\label{constt}
\eea
Equations (\ref{constf}), (\ref{consts}) and (\ref{constt}) represent the
constraints imposed on the two curvatures $\Rex$ and $\Rsex$ of the tangent
bunle and of the spinor bundle, respectively. They are nonlinear equations
and, as in the particular case previously studied in \cite{viv2}, they admitt
as solutions manifolds with both curvatures vanishing
\be
\Rex = \Rsex = 0. \label{solutions}
\ee
Let us notice that in the case when Dirac eigenspinors are subject to the
parallel transport, these equations reduce to the ones in \cite{viv2}.
However, the fact that Dirac eigenspinors are no longer subject to the
parallel transport has some nontrivial consequences which is in place to be
discussed here.

Let us assume that even if the Dirac eigenspinors change when they are
transported along the two paths, the result of the transport is still a
Dirac eigenspinor at the new point. In other words, we assume that Dirac
eigenspinors do not drop out the set of eigenspinors after parallel transport
from one point to another, which, by itself, is not an obvious thing in
the general case because it would imply that the Dirac operator defines
some global sections of the spinor bundle on $M$ in general.
Let us now consider eq. (\ref{diraceq}) transported along
the two paths and apply eq.(\ref{relhold}) for each term . We obtain the
following equation
\bea
D(\Rsex \chi^n ) - \l^n (\Rsex \chi^n ) +
i\g^a \{ (\Rex e^{\m}_{a}\6_{\m}) + \nonumber\\
(e^{\m}_a \overcirc{\o}^{(5)}_{\m bc}
(\Rex ) + (\Rex e^{\m}_a )\overcirc{\o}_{\m bc} +
e^{\m}_a K^{(5)}_{\m bc}(\Rsex ) + (\Rex e^{\m}_a
K_{\m bc}) \ss^{bc} \} = 0, \label{modchi}
\eea
where
\bea
\overcirc{\o}^{(5)}_{\n bc1NS}=
(\ne\nx e^{\m}_{a})\6_{[\n }e_{b\m ]} + \cdots
+\f{1}{2}(\ne\nx e^{\r}_{a})e^{\ss}_{b}\6_{\ss}e_{\r c}e^{c}_{\n} +
\f{1}{2}(\nx\ne e^{\r}_{a})e^{\ss}_{b}\6_{\ss}e_{\r c}e^{c}_{\n} + \cdots
\nonumber\\
K^{(5)}_{\n ab1NS}= 
\f{\i}{8}\sum_{(\n ,b,c)}(\bar{\psi}_{\n}\g_b (\nse \nsx \psi_c )
+ (\nse \nsx \bar{\psi}_{\n})\g_b \psi_c ).
\eea
Here, $\6_{[\n}e_{b \m ]}$ refers to antisymmetrization with respect to
the indices $\n$ and $\m$ only and the sums on fermion products mean
\be
\sum_{(\n ,b,c)}[A_{\n}\g_b B_c ] =
A_{\n}\g_b B_c - A_{\n}\g_c B_b + A_{b}\g_{\n} B_c .
\ee
Also, the index $NS$ indicates the nonsymmetric part 
in $\nabla_{\h}$ and $\nabla_{\x}$ and similarly for spinors with the
appropriate spin connection.

Equation (\ref{modchi}) expresses a relationship between the geometrical
structure of $M$, represented by the two curvatures, and the eigenspinors of
$D$. It represents a particular case of the previous assumption which
in its most general form should state that an eigenspinor which does not leave
the set of eigenspinors, can become a linear combination of them. In this case
we should write
\be
\chi^n \stackrel{\bf i}{\rightarrow} \chi^{n}_{\bf i}
= \sum_{k}c^{n}_{k {\bf i}}\chi^{k}_{\bf i} \label{eigspn}
\ee
where ${\bf i} =1,2$ is the index for the two paths and $c^{n}_{k {\bf i}}$
are real coefficients which may differ for the two paths. The equation
that describes this situation is
\bea
\sum_{k} ((e^{\m \bar{\nabla}_{\eta}}e^{\l \bar{\nabla}_{\xi}}D)
(e^{\m \nabla^{S}_{\eta}}e^{\l \nabla^{S}_{\xi}}\chi^k )c^{n}_{2 k}
- (e^{\m \bar{\nabla}_{\eta}}e^{\l \bar{\nabla}_{\xi}}D)
(e^{\m \nabla^{S}_{\eta}}e^{\l \nabla^{S}_{\xi}}\chi^k )c^{n}_{1 k})
\nonumber\\ 
= \l^n \sum_k (c^{n}_{2 k}(e^{\m \nabla^{S}_{\eta}}e^{\l \nabla^{S}_{\xi}}
-c^{n}_{1 k}(e^{\m \nabla^{S}_{\eta}}e^{\l \nabla^{S}_{\xi}})\chi^k ),
\label{generalf}
\eea
where $\bar{\nabla}$ stands for either $\nabla$ or $\nabla^S $ according
to the type of the object entering $D$ to which is applied.

Notice that in Eq. (\ref{generalf}) the terms in different
powers of $\m$ and $\l$ are present and the situation remains the same
for higher order terms. That shows that we cannot obtain a relationship among
the two curvatures and the eigenspinors only, because the connections in the
tangent bundle and the spinor bundle should be constrained, too.

Let us go on and see what happens when the secondaries are transported
along path 1 and path 2. In this case we should compute the
transformation of the terms entering (\ref{firstcc}), (\ref{secondcc}) and
(\ref{thirdcc}).
We should also keep in mind that ${ \chi }^{n}$ transforms, too.

In order to compute the constraint on the spacetime manifold $M$ that arises
from the first secondary,  it is worth noting that (\ref{firstcc}) already
encodes the behaviour of the eigenspinors under a diffeomorphism. Therefore
we should consider an arbitrary infinitesimal vector field ${ \z}$ defined
on $M$ with respect to which a general diffeomorphism is defined. Then $\z$
replaces $\x$ in (\ref{firstcc}). Now since $\z$ is defined everywhere
on $M$, we can consider the diffeomorphisms generated by $\z$ around the
points $Q, P, R, S$. Since, in general, $\z$ does not have to be subject
under the parallel transport, it changes when it is transported along
path 1 and along path 2 and this change should be taken into account when
the first secondary is transported along the two paths.
After performing some tedious algebra we obtain the following result
\bea
i\g^{a}\{[\Rex \z ]e^{\m}_{a}+\z [\Rex e^{\m}_{a} ]-
\Rex (e^{\n}_{a}\6 _{\n}\z^{\m})-
\nonumber \\
-2[\Rex (e^{\n}_{a}\z ^{\m}\o _{\n bc})]\ss ^{bc}\} \6 _{\m}\chi^{n}+
\nonumber \\
+b^{\m}(\z)[\Rex \6 _{\m})]\chi^{n}+b(\z)[\Rsex \chi^{n}]+
\nonumber \\
+\{ [\Rex (\l ^{n}\z ^{\m})]-D[\Rex \z ^{\m}]-
\nonumber \\
-i\g ^{a}[\Rex (e_{a})+
(\Rex e^{\n}_{a}) \o _{\n bc} \ss^{bc} +
\nonumber \\
+[e^{\n}_{a} (\overcirc{\o}_{\n bc}(\Rex )+
{K}_{\n bc}(\Rsex ))]\ss ^{bc}] \z ^{\m}\} \6 _{\m}\chi^{n} +
\nonumber \\
+c(\l ,\z )^{\m}(\Rex \6 _{\m}) \chi^{n} +
c(\l ,\z )^{\m} \6 _{\m} [\Rsex \chi^{n}]+
\nonumber \\
i\g^{a}\{ [\Rex \z ](e^{\m}_{a} \o _{\m bc} )+
\z [\Rex (e^{\m}_{a} \o _{\m bc})] \ss ^{bc}\}  \chi^{n}+
\nonumber \\
+ f(\z) [\Rsex \chi^{n}] = 0,
\label{finalcc}
\eea
where $\o _{\m bc} = \overcirc{\o}_{\m bc} + K_{\m bc}$.

We are going to compute now the transport of the second secondary.
Here, we must pay attention to some other subtleties. Namely, when
(\ref{secondcc}) is transported along any of the two paths,
the parameter $\th_{ab} = -\th_{ba}$ can change, because this is the
parameter of the local $SO(4)$ group. Similarly, $M_{ab}$ can in
principle change, too, to another $SO(4)$ matrix. We can make the
assumption which do not affect the generality of the result, that $M_{ab}$
belongs to a basis of $SO(4)$ and that this basis do not change. Then we
can write:
\be
M_{ab2}-M_{ab1} = \sum_{k}(d_{k2}-d_{k1})M_{ab} ,
\label{mmm}
\ee
where the sum is over all of the elements of the basis of $SO(4)$. Then
performing the same steps as in the case of the first secondary
we obtain the following result:
\bea        
(\Rex \th^{a}_{a})D \chi^{n} + \th^{a}_{a} D (\Rsex \chi^{n})+
\nonumber \\
i\th^{a}_{a} \g^{d}
[(\Rex e_{a})+ [\overcirc{\o}^{(5)}_{\n bc}(\Rex ) +
{K}^{(5)}_{\n bc}(\Rsex )]\ss ^{bc}] \chi^{n}+
\nonumber \\
\{ \g ^{c} [\Rex e_{c}^{\m} ] ([\th \ss , {\o}_{\m de} ] \ss ^{de})+ 
\g ^{c} e^{\m}_{c}([\Rex \th \ss , {\o}_{\m de}] \ss ^{de}-
\nonumber \\
\g ^{c} \sum_{k=1}^{4}(d_{2k}e^{\m \ne \l \nx }-d_{1k}e^{\l \nx \m \ne })
(\6_{\m} \th \ss M^{j})\} 
+g(\th )[\Rsex \chi^{n}]+
\nonumber \\
i\{ [\Rex (\l ^{n}\th \ss )-D [\Rex \th \ss ]-i\g^{a} [(\Rex e_{a}) +
\nonumber \\
(\Rex e^{\n }_{a}) \o_{\n bc} +
e^{\n }_{a}(\overcirc{\o}_{\n bc}(\Rex ) +
{K}_{\n bc}(\Rsex ))] \ss ^{bc} \th \ss\} \chi^{n} +
\nonumber \\
i(\l ^{n}-D)[\Rex \th \ss ]  \chi^{n} +h(\th ) [\Rsex \chi^{n}] = 0.
\label{finalccc}
\eea
We can now work out the constraint that arises from the transport of the
third secondary. Proceeding along the same line as in the previous
case, after somewhat simpler computations, we obtain the following equation
\bea
\frac{1}{2} \g ^{a} [\Rsex (\bar{\e }\psi ^{\m})] \6 _{\m} \chi^{n} +
j^{\m }_{a} [\Rex \6 _{\m}]\chi^{n} +
j^{\m }_{a} \6 _{\m}[\Rsex \chi^{n} ]+
\nonumber \\
\frac{1}{2} \g ^{a} [\Rsex (\bar{\e }\psi ^{\m})]
\o _{\m cd}\ss ^{cd} \chi^{n} +
\frac{1}{2} \g ^{a} \bar{\e }\psi ^{\m}
[\overcirc{\o}^{(5)}_{\m cd}(\Rex )+
{K}^{(5)}_{\m cd}(\Rsex )]\ss ^{cd}\chi^{n} +
\nonumber \\
K_{a}[\Rsex \chi ^{n}] +
\{ [\Rex e_{a}^{\m}]A_{\m cd}\ss ^{cd}+
e_{a}^{\m} \{ [\Rex (e_{\m}^{f} e_{\rho c}e_{\ss d})]A_f^{\rho \ss}+
\nonumber \\
(e_{\m}^f e_{\rho c}e_{\ss d}) \{ [\Rsex \bar{\e } ]
\g_{5}\g_{f}D_{\l }\psi_{\th }\e ^{\ss \rho \l \th }+
\nonumber \\
\bar{\e } \g_{5}\g_{f} \{ [\Rex \6_{\l }] \psi_{\th } +
 \6_{\l } [\Rsex \psi_{\th }] +
\nonumber \\
\frac{1}{2} [\overcirc{\o}^{(5)}_{\l mn}(\Rex )\ss^{mn}\psi_{\th }+
\overcirc{\o}^{(0)}_{\l mn} \ss^{mn} (\Rsex \psi_{\th })]+
\nonumber \\
\frac{1}{8} [\sum_{(\l ,m,n)} \bar{\psi }_{\l}\g_{m} {\psi }_{n}\ss^{mn}
(\Rsex \psi_{\th }) +
\sum_{(\l ,m,n)}[\Rsex (\bar{\psi }_{\l}\g_{m} {\psi }_{n})]
\ss^{mn}\psi_{\th } \} \e^{\ss \rho \l \th }\} \ss^{cd}-
\nonumber \\
\frac{1}{2} [\Rex e_{\m d}]A^{e}_{ec} \ss^{cd}-
-\frac{1}{2} e_{\m d} \{ [\Rex (e_{\pi e}e_{\chi c})]A^{e\pi \chi }+
\nonumber \\
e_{\pi e}e_{\chi c} \{ [\Rsex \bar{\e }]\g_{5}\g^{e}
D_{\l }\psi_{\rho }\e ^{\chi \pi \l \rho }+
\bar{\e } \g_{5}\g^{e}
\{ [\Rex \6 _{\l }] \psi_{\rho } +
\nonumber \\
\6 _{\l }[\Rsex \psi_{\rho }] +
\frac{1}{2} [\overcirc{\o}^{(5)}_{\l mn}(\Rex )\ss^{mn}\psi_{\rho }+
\overcirc{\o}^{(0)}_{\l mn} \ss^{mn} (\Rsex \psi_{\rho })]+
\nonumber \\
\frac{1}{8} [\sum_{(\l ,m,n)} \bar{\psi }_{\l}\g_{m} {\psi }_{n}\ss^{mn}
(\Rsex \psi_{\rho }) +
\nonumber \\
\sum_{(\l ,m,n)}[\Rsex (\bar{\psi }_{\l}\g_{m} {\psi }_{n})]
\ss^{mn}\psi_{\rho } \} \e^{\chi \pi \l \rho }\} \} \ss^{cd}+
\nonumber \\
+\frac{1}{2} (-\rightarrow +, d \rightarrow c)\} \} \chi^{n}+
+l_{a}[\Rsex\chi^n] = 0,
\label{finalcccc}
\eea
where   $\frac{1}{2} (-\rightarrow +, d \rightarrow c)$ means
that the terms with   $\frac{1}{2}$ change sign and the indices $d$ and $c$
get interchanged in these terms. Some remarks regarding our notations are
in order here. For all of the equations (\ref{finalcc}), (\ref{finalccc})
and (\ref{finalcccc}) $R$ acts on the product $AB$ 
according to the Leibniz rule.
The same for $R^{S}$, with no change of sign. 
$\overcirc{\o}^{(5)}_{\m bc} (\Rex)$
and ${K}^{(5)}_{\m bc}(\Rsex )$ mean that only the nonsymmetric part of
$\overcirc{\o}^{(5)}_{\m bc}$ and ${K}^{(5)}_{\m bc}$ in 
$\ne \nx$ are considered
and that these parts, by substracting the expression for the first path
from the expression for the second path, 
depend on the corresponding curvatures.
$\th \ss $ stands for $\th_{ab} \ss^{ab} $ and $M^k = M^k_{cd}\ss ^{cd}$.

We cannot comment here on the general solution of equations (\ref{constf}),
(\ref{consts}), (\ref{constt}), (\ref{generalf}), (\ref{finalcc}),
(\ref{finalccc}) and (\ref{finalcccc}). These equations are nonlinear
and highly nontrivial. We see that apparently they do not admitt as a
solution the spacetime manifolds with both curvatures of the tangent
bundle and of the spinor bundle vanishing, as was the case studied
in \cite{viv2}. The main obstructions for that seem to
be the equations (\ref{generalf}) and (\ref{finalccc}). Should we have impose
that "totaly flat" spacetime satisfy all of the equations, we would have
easily checked out that (\ref{generalf}) and (\ref{finalccc}) reduce
to the following equations
\bea
\sum_{k}(c^{n}_{2k}-c^{n}_{1k})
(e^{\m \ne}e^{\l \nx}D)
(e^{\m \nse}e^{\l \nsx}\chi ^k) =
\nonumber \\
= \l ^{n} \sum_{k}(c^{n}_{2k}-c^{n}_{1k})
(e^{\m \nse}e^{\l \nsx}\chi ^k)
\label{partica}
\eea
and
\be
\g^{c}\sum^{4}_{\a = 1}[
(d_{2\a }-d_{1\a })
e^{\m \ne}e^{\l \nx}
(e^{\m }_{c}\6_{\m }\th \ss M^{\a })
]\chi ^n) = 0.
\label{particb}
\ee
Thus we can see that even if the two curvatures vanish, there remain
two equations that must be satisfied by $\chi ^k$'s and the two connections.
They are automatically satisfied for $c^{n}_{2k}=c^{n}_{1k}$ and
$d_{2\a }=d_{1\a }$, that is when the gauge transformations do not mixt
either the eigenspinrs or the matrices $M$. The first assumption is almost
trivial since we do not expect that diffeomorphisms, local rotations or
local supersymmetry exchange the spinors corresponding to different Dirac
eigenvalues, e.g. to different masses. The second assumption is again
natural once we do not expect that $SO(4)$ matrices be affected by
diffeomorphisms, i.e. by the parallel transport.

The conclusion is that, with these two natural assumptions at hand, the
manifolds with vanishing tangent curvature and spinor curvature, admitt Dirac
eigenvalues as observables in the most general case.

\section{Summary and Concluding Remarks}
In this paper we have derived the most general set of constraints 
on the curvature of tangent bundle and the curvature of spinor bundle,
respectively, under which the spacetime manifold admitts Dirac eigenvalues 
as observables of Euclidean supergravity. A solution of these constraints 
has been shown to be given by spacetime manifolds with both curvatures vanishing. 
If Dirac eigenspinors remain in the set of eigenspinorsafter being subject 
to parallel transport, the connections in the tangent bundle are constrainted, too.

The equations derived above are simpler if the spacetime has a vanishing 
bosonic torsion.
In this case, owing to Ricci identity, the covariant derivatives of the
functions on $M$ that appear in these equations, as $\th _{ab}$, commute,
but some attention should be paid to the components of different tensor
and spinor objects. However, it is not obvious at this stage that the
bosonic torsion should vanish. Usually, the constraints on the supertorsion
appear in a natural way in the superfield approaches to supergravity, when some
symmetries are imposed on the action. However, this is not the case here
and finding the constraints on the torsions of $\nn $ and $\nabla^{S}$ is an
interesting problem.

\acknowledgements
C. Ciuhu would like to thank R. Daemi for hospitality at Abdus Salam
International Centre for Theoretical Physics and I.V.Vancea would like
to thank A. van Proeyen and W. Troost for discussions.

\end{document}